\documentstyle[prl,aps,graphicx]{revtex}
\newcommand{\mra}  {\rightarrow}
\begin{document}
\draft
\title{\bf Microcanonical Thermodynamics of\\ First Order Phase Transitions\\
studied in the Potts Model }
\author{ D.H.E. Gross, A. Ecker and X.Z. Zhang $^*$}
\address{ Hahn-Meitner-Institut
Berlin, Bereich Theoretische Physik, \\14109 Berlin, Germany\\
$^*$ Institute of Atomic Energy, Beijing, China}
\date{\today}
\maketitle
\begin{abstract}
Phase transitions of first and second order can easily be distinguished in
small systems in the {\em microcanonical} ensemble.  Configurations of phase
coexistence, which are suppressed in the conventional canonical formulation,
carry important information about the main characteristics of first order phase
transitions like the transition temperature, the latent heat, and the
interphase surface tension. The characteristic backbending of the
microcanonical caloric equation of state $T(E)$ (not to be confused with the
well known Van der Waals loops in ordinary thermodynamics or mean field
approximations) leading to a negative specific heat is intimately linked to the
interphase surface entropy.
\end{abstract}
\pacs{Keywords: Microcanonical thermodynamics, phase transitions,
surface tension.}

Microcanonical thermodynamics describes the depen\-dence of the volume
$\Omega_N$ of the N-body phase\--space of an interacting many\--body system on
the globally conserved quantities like the total energy, momentum, angular
momentum, mass, charge etc.. This is the fundamental starting point of any
statistical definition of thermodynamics. By Laplace transform of $\Omega_N(E)$
from the extensive quantities like the energy to intensive ones like inverse
temperature ($\beta=1/T$) one obtains the more familiar Gibb's canonical
partition function $Z(\beta)$.  For systems interacting by short range two-body
forces (with hard cores) both formulations are identical in the thermodynamic
limit of infinitely many particles at the same particle density 
\cite{vanhove49}.  However, the two formulations are different for finite
systems and more essentially are different even in their physical content for
systems with long range forces like gravity
\cite{thirring70} or Coulomb dominated systems see for example
\cite{gross95}.  This will be discussed in a forthcoming paper.

One of the most interesting phenomena in thermodynamics are phase transitions.
Computer simulations of simple models give a deep insight into the
mechanism.  Naturally these calculations can only be performed for small
systems. The main characteristics like the transition temperature $T_{tr}$, the
specific latent heat $q_{lat}$, and the interphase surface tension
$\sigma_{surf}$ are extrapolated to the infinite system. In the thermodynamic
limit it should not matter whether the calculations for the finite systems are
performed canonically or microcanonically. However, we will show for the
two-dimensional, $q=10$ states, Potts model, which has a clear phase transition
of first order from ordered to disordered spins, the extrapolations converge
faster when started from the microcanonical-ensemble. Moreover, our
investigation will clarify some of the most striking features of finite
micro-ensembles, the occurence of the S-bend in the microcanonical caloric
equation of state $T(E)$ and the occurence of a negative specific heat at a
phase transition of first order, see e.g.:
refs.\cite{gross95,gross141,lyndenbell95,lyndenbell95a,doye95,wales95}. \{Here
we do not discuss the negative specific heat in thermodynamically unstable
systems with long-range interaction as were discussed by Thirring
\cite{thirring70}. The systems we discuss here are thermodynamically stable
with short-range interactions. They behave properly in the canonical ensemble
and have a positive specific heat there and in the thermodynamic limit.\}

The Hamiltonian of the Potts model is defined as
\begin{equation}
H=\sum_{i<j}{'\{1-\delta_{\sigma_i,\sigma_j}\}}.
\end{equation}
on a two dimensional lattice (here with periodic boundary conditions) of
$N=L*L$ spins with $q=10$ possible values (components).
The sum is over pairs of nearest neighbour lattice points only and $\sigma_i$
is the spin state at the i-th lattice point. For a microcanonical ensemble we
have to calculate the partition sum over all possible different configurations
$\nu$ with the same total energy $E$ :
\begin{equation}
\Omega_N(E)=\sum_{\nu}{\delta_{E_{\nu},E}} \;\;.
\end{equation}
This is done by using the ''Ergodic Microcanonical Metropolis Monte Carlo
Algorithm'' ($M\!M\!M\!C$) \cite{lee95,gross148}. [The main idea here is to
sample the energy in a narrow band $E-4\le E_i \le E+4$ by Metropolis sampling.
From the slope of $ln(P(E_i))|_{E}$, in this energy-band one can determine
$\beta_{micro}(E)$]. $\Omega_N(E)$,
the specific entropy $s(\epsilon)$, and the thermodynamic temperature are
related 
by (we put Boltzmann's constant $k=1$) :\begin{eqnarray} N
s(\epsilon)=ln\{\Omega_N(E=N*\epsilon)\}\\
\beta_{micro}(\epsilon)=\frac{\partial s(\epsilon)}{\partial \epsilon}\\
T(\epsilon)=\frac{1}{\beta_{micro}(\epsilon)}.
\end{eqnarray}
The connection to the canonical (Gibbs-) free energy $F(\beta)$ is via the
Laplace transform in saddle-point approximation:
\begin{eqnarray}
Z(\beta)=\int_0^{\infty}{\Omega_N(N\epsilon)\;e^{-\beta N
\epsilon}\;Nd\epsilon}
\label{laplace}\\ 
\sim e^{N[s(\bar{\epsilon})-\beta\bar{\epsilon}]}\;T\sqrt{2\pi
N/c(\bar{\epsilon})}\\ 
\frac{\partial}{\partial\epsilon}s(\epsilon)|_{\bar{\epsilon}}=\beta
\label{station}\\ 
F(\beta)=-T*ln\{Z(\beta)\}\\
c(\bar{\epsilon})=-\left.\frac{\partial ^2s}{\partial\epsilon ^2} 
\frac{1}{\left(\frac{\partial s}{\partial \epsilon}\right)^2}\right|_{\varepsilon=\bar{\varepsilon}}.
\end{eqnarray}

Plotting $s(\epsilon)$ and $\beta_{micro}(\epsilon)$ v.s. $\epsilon$ we see a
convex intruder in $s(\epsilon)$ (fig.1), where $s(\epsilon)$ is reduced
compared to its concave hull, the straight tangent line \{$\sim$ the canonical
$s(\epsilon)$\} by $\Delta s_{surf}$. For an infinite system van Hove's
theorem \cite{vanhove49,katsura50,katsura54,hill55} forbids the existence of
any convex part in $s(\epsilon)$, because then the system gains entropy if it
would divide spontaneously into two equal pieces one with ordered spins and
with entropy $s_1=s(\epsilon_1)$, the other with disordered spins and entropy
$s_3=s(\epsilon_3)$. Both pieces together would have the {\em larger} entropy
$\bar{s}=(s_1+s_3)/2\ge s_2=s(\epsilon_2=(\epsilon_1+\epsilon_3)/2)$. This is
very nicely discussed by H\"uller in several recent articles e.g.
\cite{hueller94a}.

This argument does not apply to a finite system as the new surface of
phase-separation reduces the specific entropy by $\Delta s_{ph-sep}$ which is
proportional to the number of lattice spins being fixed in the dividing surface
i.e. $\Delta s_{ph-sep}\propto L/N \propto L^{-1}$. With growing size of the
system, $L\mra\infty$,  $\beta_{micro}(\epsilon)$ approaches the horizontal
line $\beta_{can}(\epsilon)\equiv \beta_{tr}$, determined by the Maxwell
construction of equal areas ${\cal A}_l={\cal A}_r$ on the left side below
$\beta_{tr}$ and on the right above it, lower half of Fig.1..  On the other
hand the surface entropy $\Delta s_{surf}$ is equal to one of these areas $\cal
A$, half of the shaded area  under the S-like oscillation of
$\beta_{micro}(\epsilon)$ between the curve and the line $\beta_{tr}$ .
$\beta_{micro}(\epsilon)$ can directly be calculated by $M\!M\!M\!C$ so also
the area ${\cal A}=\Delta s_{surf}$. As the Potts model has only nearest
neighbour interactions the interphase surface tension $\sigma_{surf}=T*\Delta
s_{surf}$ consists of an entropy part only. As the entropy of the special
configuration with planar phase-separation can only be $\le s(\epsilon)$ we
have $\Delta s_{surf} \le \Delta s_{ph-sep}=O(1/L)$.  We make the
\underline{conjecture} that {\em the area $N*\cal A$ under $N*s(\epsilon)$
scales like the cross-section of the lattice ($\propto L$).}

The intimate link of ${\cal A} = \Delta s_{surf}$ to the surface tension can
also be seen from the way how the surface tension is calculated in conventional
canonical Monte Carlo simulations e.g.\cite{binder82,borgs92,janke94}: Here the
probability $P_T(\epsilon)$ to find the system with temperature $T$ at an
energy $\epsilon$ shows at the transition temperature two well separated maxima
with a deep valley in between.  The ratio of the minimum to the maximum is
by \cite{binder82} equal to:
\begin{equation}
N*\Delta s_{surf}=-ln{\{\frac{P_{tr}(min)}{\sqrt{P_{tr}(max1)P_{tr}(max2)}}\}}
\end{equation}

Fig. 2 shows $L*\Delta s_{surf}$ vs. the inverse lattice size $1/L$ in
comparison with the surface tension determined by the multi-canonical method
\cite{billoire93}. (Actually the quantity shown is twice the surface tension as
we use periodic boundary conditions and consequently two cuts have to be made
to separate the two phases.) Both results  scale to the analytically known
asymptotic limit \cite{borgs92,billoire93}.  We make an interesting observation
here : The microcanonical surface tension depends much less on the size of the
lattice than the canonical one. That means one may get its asymptotic value
already at {\em smaller} sizes of the system than canonically.

Insight into the possible mechanism leading to this weaker finite size scaling
is obtained from the projected correlation function $g^{(1)}(dx)$
\cite{janke95}.

\begin{equation}
g^{(1)}(dx=|i_x-j_x|)=\frac{1}{L}\sum_{i_y,j_y}{\{\delta_{s_i,s_j}-
\frac{1}{q}\}}
cos\{\frac{2\pi}{L}(i_y-j_y)\}.
\end{equation}
In the canonical ensemble it approaches 
\begin{eqnarray}
g^{(1)}(dx)\propto cosh\{(dx-L/2)/\xi^{(1)}\}\\
\xi^{(1)}=\xi/\sqrt{1+(2\pi\xi/L)^2}
\end{eqnarray}
with $\xi$ the correlation length of the infinite lattice \cite{janke95}.
Figure 3 shows $g^{(1)}(dx)$ at an excitation energy $\epsilon=0.7$ in the
middle of the coexistence region.  In the microcanonical ensemble $g^{(1)}$
does not need to be an exponential-like function and in fact it drops down 
faster in the region of coexistence ($0.33\le\epsilon\le1.05$) . In any case,
the region of coexistence of both phases, gets suppressed for large $N$ by the
Laplace transform eq.(\ref{laplace}). It does not contribute because in this
region there is no stationary point, no solution of the equation
(\ref{station}) and all information contained in events showing the coexistence
of both phases is lost in the canonical ensemble. This will more thoroughly
be investigated in a forthcoming publication.

Most of these calculations were done on different workstations. For the
largest lattice ($L=100$) we performed $3.2*10^6$ sweeps. To destroy
correlations due to the previous energy step the spins in an arbitrarily chosen
rectangular part of the lattice were first uniformly lifted one value higher up
(cyclic) and the energy was restored by switching random spins of the lattice.
 This procedure
has the advantage that the perturbation per latice point is $\propto 1/L$ and
all correlations inside the rectangle are conserved.  In between these
macroscopic moves the spins were individually updated in sequence one after the
other. As can be seen from fig.1 the accuracy for $L=100$ is not optimal but
sufficient for the demonstration of the method. Due to bad computational
conditions here at the HMI it was not yet possible to do better.

An important lesson can be learned here : Following M. Fisher
\cite{fisher67} a phase transition of first order is distinguished from a
transition of second order by a non-vanishing interphase surface tension. That
means {\em the microcanonical caloric equation of state for a phase transition
of first order must show an S-oscillation and consequently a negative specific
heat $c=\partial \epsilon/\partial T$}. This solves the outstanding problem of
the backbending and of the negative specific heat at a phase transition of
first order in the microcanonical ensembles discussed before by many authors
e.g.  ref.\cite{thirring70,lyndenbell95a,doye95,wales95}. It also appears at
the condensation phase transition of selfgraviting astrophysical bodies which
again must be treated {\em microcanonical} as the long range of gravity does
not allow for the transition to the thermodynamic limit
\cite{thirring70}.
\{A selfgraviting system at constant energy behaves microcanonically quite
differently from the same system at constant temperature (canonical). Whereas
in the second case it can totally implode at low temperatures and convert all
surplus energy to the heatbath, in the first case it can only partly implode
because all energy must be stored in the noncollapsed rest of the system.\}
 
Figure 4 shows the three crossing points $\epsilon_1,\epsilon_2,\epsilon_3$ of
the Maxwell line $\beta_{tr}$ with $\beta_{micro}(\epsilon)$ as function of the
inverse lattice length $1/L$.  The difference $q_{lat}=\epsilon_3-\epsilon_1$
at $L=\infty$ is the specific latent heat of the transition, which is
analytically known from Baxter \cite{baxter73}.  Our results up to the lattice
length $L=100$ agree well when extrapolated linearly in $1/L$ to $L=\infty$.
Only the middle solution of $\beta_{micro}(\epsilon_2)=\beta_{tr}$ fails.
However, this point is numerically as well as also physically the instability
point. For the latent heat it is irrelevant.  Again the finite size scaling of
the microcanonical $\epsilon_i$ turns out to be very weak. It is rather
surprising to find this simple scaling at such small lattices. As the theory of
finite-size scaling for the microcanonical ensemble does not yet exist we have
to take this promising result with care. However, as was discussed by H\"uller
\cite{hueller94} this
may be at no surprise, as much of the size dependence of the canonical ensemble
comes from the trivial factor $N$ in the exponent of the Laplace transform
(\ref{laplace}) which has nothing to do with the physics contained in the
specific entropy $s(\epsilon)$.
 In a future publication we will discuss
possible reasons for this weak sensitivity of the microcanonical Potts model to
the finite size of the lattice and we will compare the microcanonical
correlation functions with the canonical ones.

We believe our findings with the $q=10$-states Potts model are characteristic
for all microcanonical ensembles. In fact the backbending of the caloric
equation of state $T(E)$ was found in many other finite system e.g. fragmenting
nuclei
\cite{gross95}, fragmenting atomic clusters \cite{gross141}, "melting" phase
transitions in van der Waals clusters \cite{wales95,lyndenbell95} . More work,
however, must be done to understand what the "surface" entropy, the area $\cal
A$ under the S-shape of $\beta_{micro}(\epsilon)$, means in these cases which
cannot be extrapolated to infinite systems. The scaling property of the
microcanonical ensemble is yet unknown and must be examined.

We want to thank the Fachbereich Physik of the Freie Universit\"at Berlin and
the Institut f\"ur Hochenergiephysik, DESY, Zeuthen for the excellent
computing conditions. Only with their generous support we were able to perform
these numerical calculations.

\includegraphics*[bb = 65 11 444 412, angle=-90, width=15cm, 
clip=true]{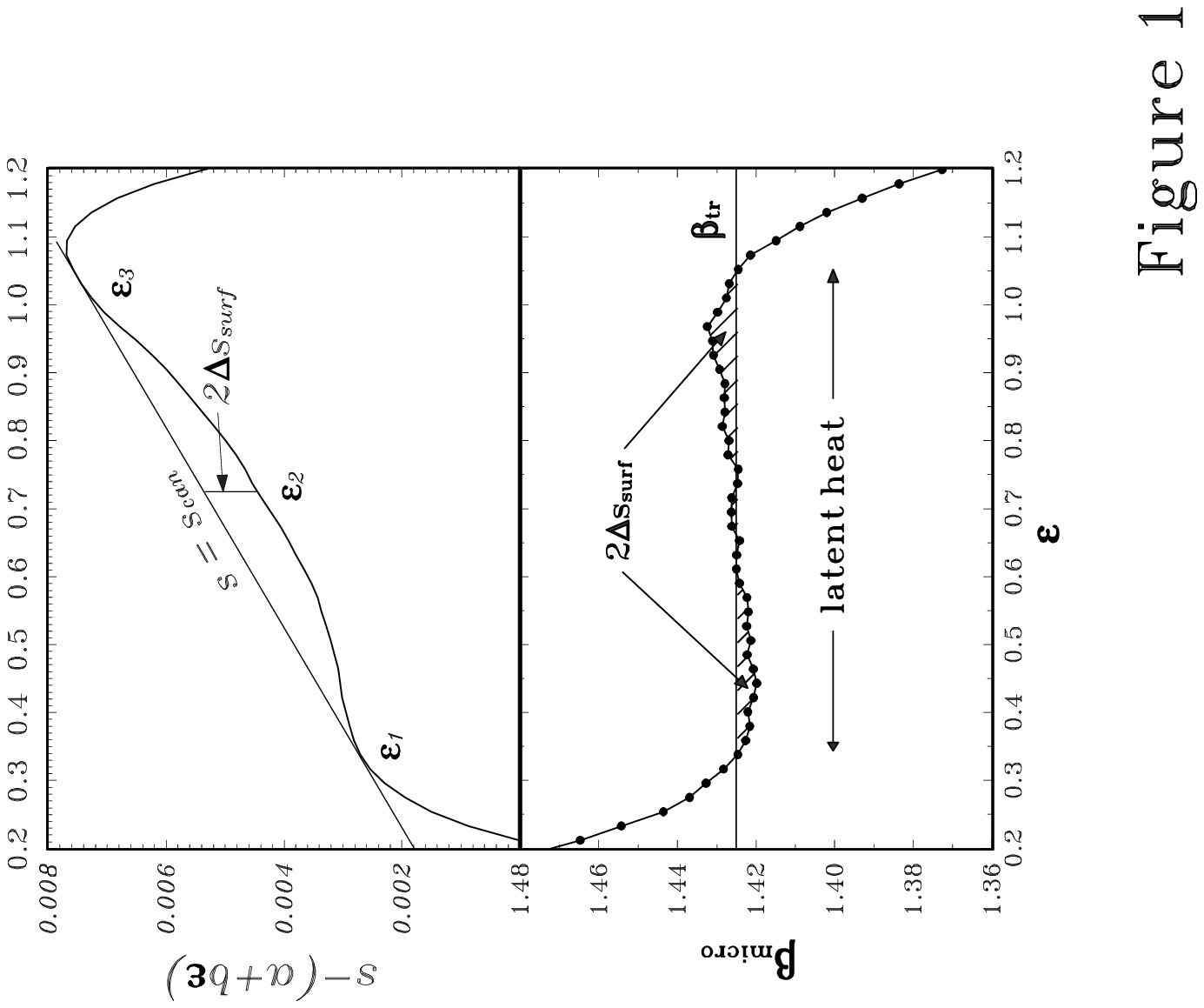}\\~\\
\begin{figure}
\caption{ a) Specific entropy
\protect{$s(\epsilon)=\int_0^{\epsilon}{\beta_{micro}(\bar{\epsilon})
d\bar{\epsilon}}$} vs.  the specific energy $\epsilon$ for the 2-dim.  Potts
model with $q=10$ on a $100*100$ lattice.  In order to visualize the anomaly of
the entropy the linear function $a+b\epsilon$ ($a=s(0.2119)$, $b=1.4185$) was
subtracted. Because we use periodic boundary conditions one needs two cuts to
separate the phases  and the convex intruder is twice the surface-entropy.
\protect\newline
b) Inverse temperature $\beta_{micro}(\epsilon)=1/T(\epsilon)$ as directly
calculated by $M\!M\!M\!C$ } 
\end{figure}
\begin{figure}
\includegraphics*[bb = 0 17 494 597, angle=-90, width=15cm, 
clip=true]{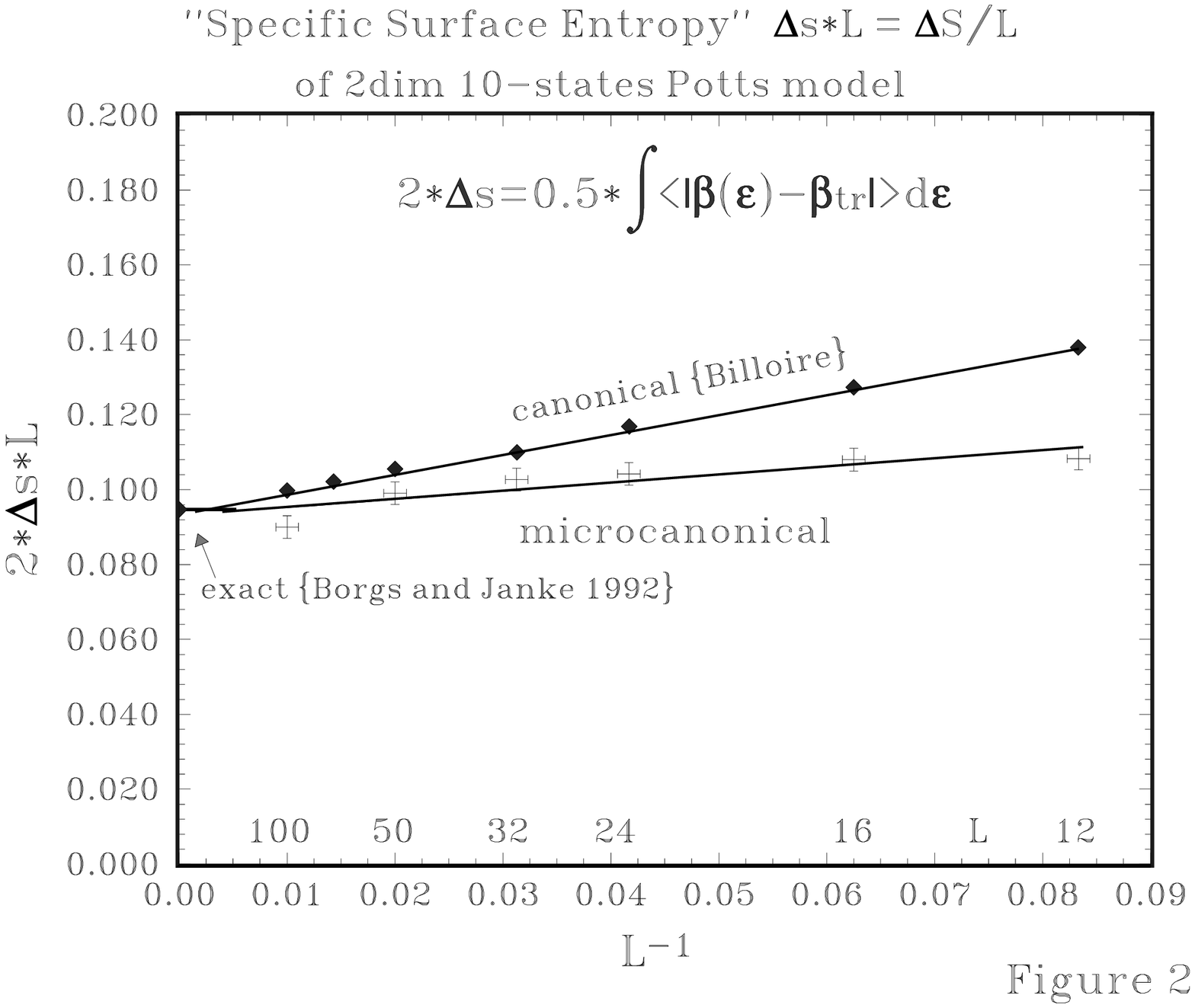}\\~\\
\caption{Surface entropy $2\Delta s_{surf}$ evaluated as half the shaded area
in figure 1 as function of the inverse lattice length $1/L$ compared to the
same quantity from the multi-canonical method of Billoire
\protect\cite{billoire93}.  }
\end{figure}
\begin{figure}
\includegraphics*[bb = 40 33 491 566, angle=-90, width=15cm, 
clip=true]{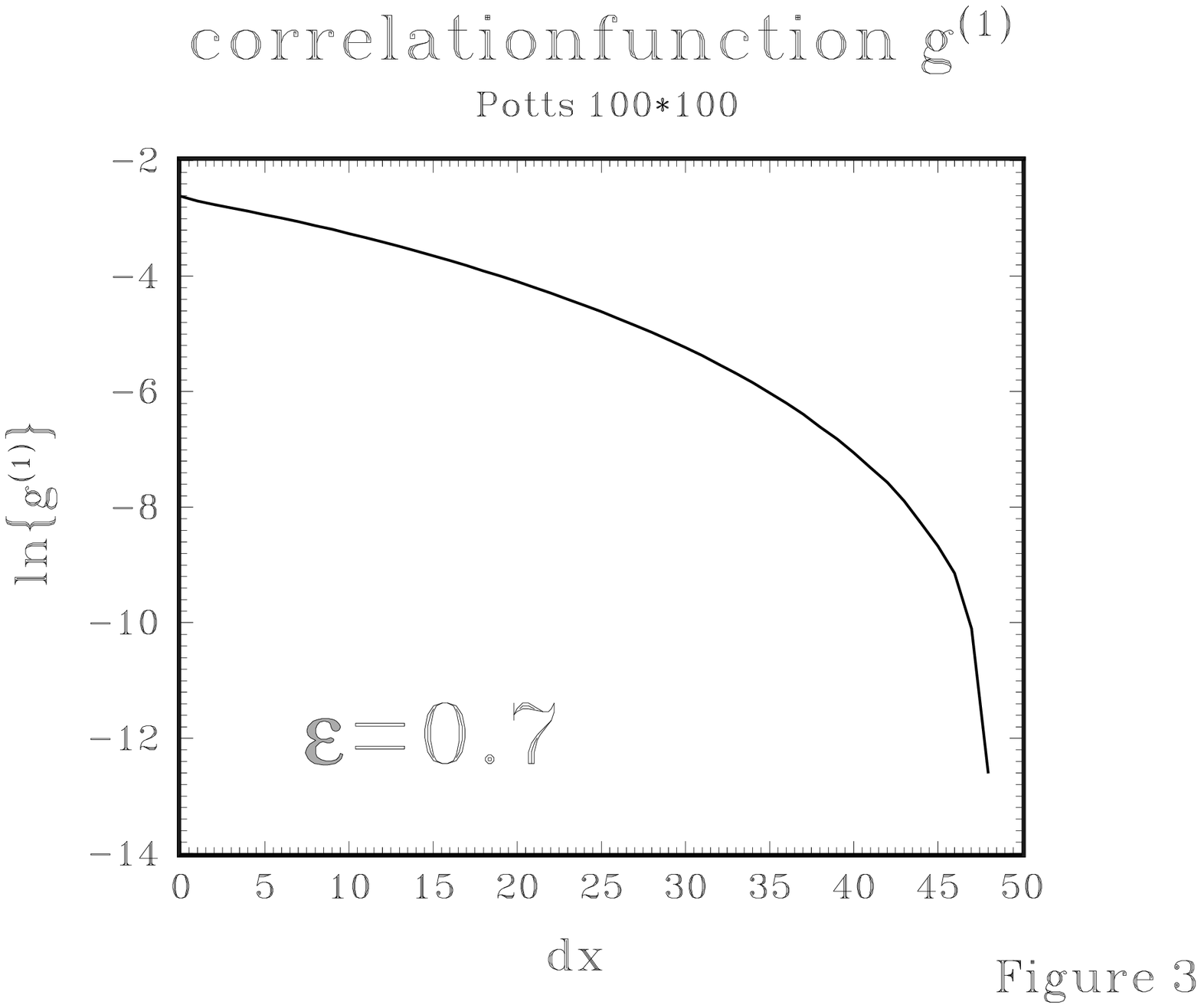}\\~\\
\caption{The microcanonical correlation function $g^{(1)}(dx)$ at
$\epsilon=0.7$ in the coexistence region.}
\end{figure}
\begin{figure}
\includegraphics*[bb = 27 70 498 630, angle=-90, width=15cm, 
clip=true]{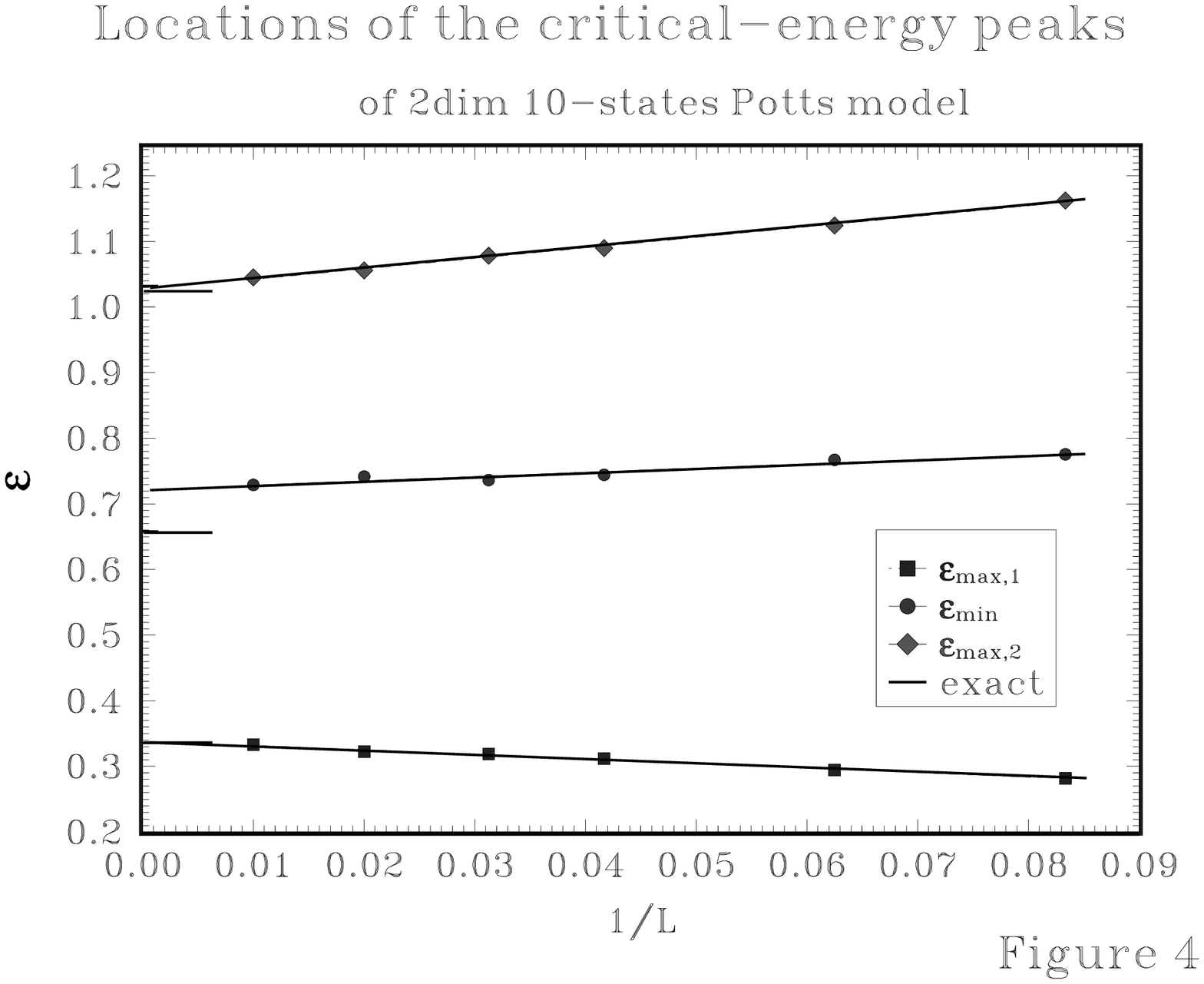}\\~\\
\caption{The three solutions of $\beta_{micro}(\epsilon_{1,2,3})=\beta_{tr}$ as
function of the inverse lattice length $1/L$. }
\end{figure}

\end{document}